\documentclass[a4paper]{spie}  

\usepackage{amsmath,amsfonts,amssymb}
\usepackage{graphicx}

\usepackage{amsmath,amssymb,amsfonts}
\DeclareMathOperator*{\argmin}{arg\,min}
\usepackage{textcomp}
\usepackage{xcolor}
\usepackage{tabularx}
\usepackage{booktabs}
\usepackage[font=small]{subfig}
\usepackage[colorlinks=true, allcolors=blue]{hyperref}
\title{A Jigsaw Puzzle Solver-based Attack on Block-wise Image Encryption for Privacy-preserving DNNs}

\author[a]{Tatsuya Chuman}
\author[a]{Hitoshi Kiya}
\affil[a]{Tokyo Metropolitan University, Tokyo, Japan}

\authorinfo{Further author information:\\Tatsuya Chuman: E-mail: chuman-tatsuya1@ed.tmu.ac.jp\\  Hitoshi Kiya: E-mail: kiya@tmu.ac.jp}

\begin{document} 
\maketitle
\begin{abstract}
   Privacy-preserving deep neural networks (DNNs) have been proposed for protecting data privacy in the cloud server.
   Although several encryption schemes for visually protection have been proposed for privacy-preserving DNNs, several attacks enable to restore visual information from encrypted images.
   On the other hand, it has been confirmed that the block-wise image encryption scheme which utilizes block and pixel shuffling is robust against several attacks.
   In this paper, we propose a jigsaw puzzle solver-based attack to restore visual information from encrypted images including block and pixel shuffling.
   In experiments, images encrypted by using the block-wise image encryption are mostly restored by using the proposed attack.
\end{abstract}


\section{Introduction}
In recent years, the rapid development of deep neural networks (DNNs) enables us to implement speech recognition and image classification with high accuracy\cite{lecun2015deeplearning}.
Various perceptual encryption methods have been proposed to generate visually-protected images\cite{nakamura_2015,Warit_IEEEtrans,AprilPyone_IEEEtrans,kiya_2022_2}.
Although information theory-based encryption (like RSA and AES) generates a ciphertext, images encrypted by the perceptual encryption methods can be directly applied to some image processing algorithms or image compression algorithms\cite{kiya2022overview}.
Numerous encryption schemes have been proposed for privacy-preserving DNNs, but several attacks including DNN-based ones were shown to restore visual information from encrypted images\cite{tanaka_ICCETW,ito_IEEEaccess,Jialong_ICCC}.
Therefore, encryption schemes that are robust against various attacks are essential for privacy-preserving DNNs.
Privacy-preserving deep neural networks (DNNs) have been proposed for protecting data privacy in the cloud server.
\par
Several perceptual encryption schemes have been proposed to generate visually protected images for privacy-preserving deep neural networks (DNNs).
However, the latest attacks succeeded in reconstructing visual information form encrypted images\cite{kiya2022overview}.
Therefore, encryption schemes need to have robustness against various attacks and compatibility with the privacy-preserving DNNs.
On the other hand, the use of the vision transformer (ViT) enables image encryption for DNNs to apply block scrambling and pixel shuffling which enhance robustness against several attacks\cite{AprilPyone_compress,kiya_2022,qi_vit_2022}. 
In this paper, we evaluate the security of the block-wise image encryption combined with the ViT.

\begin{figure}[t]
	\begin{center}
	\includegraphics[width=9cm]{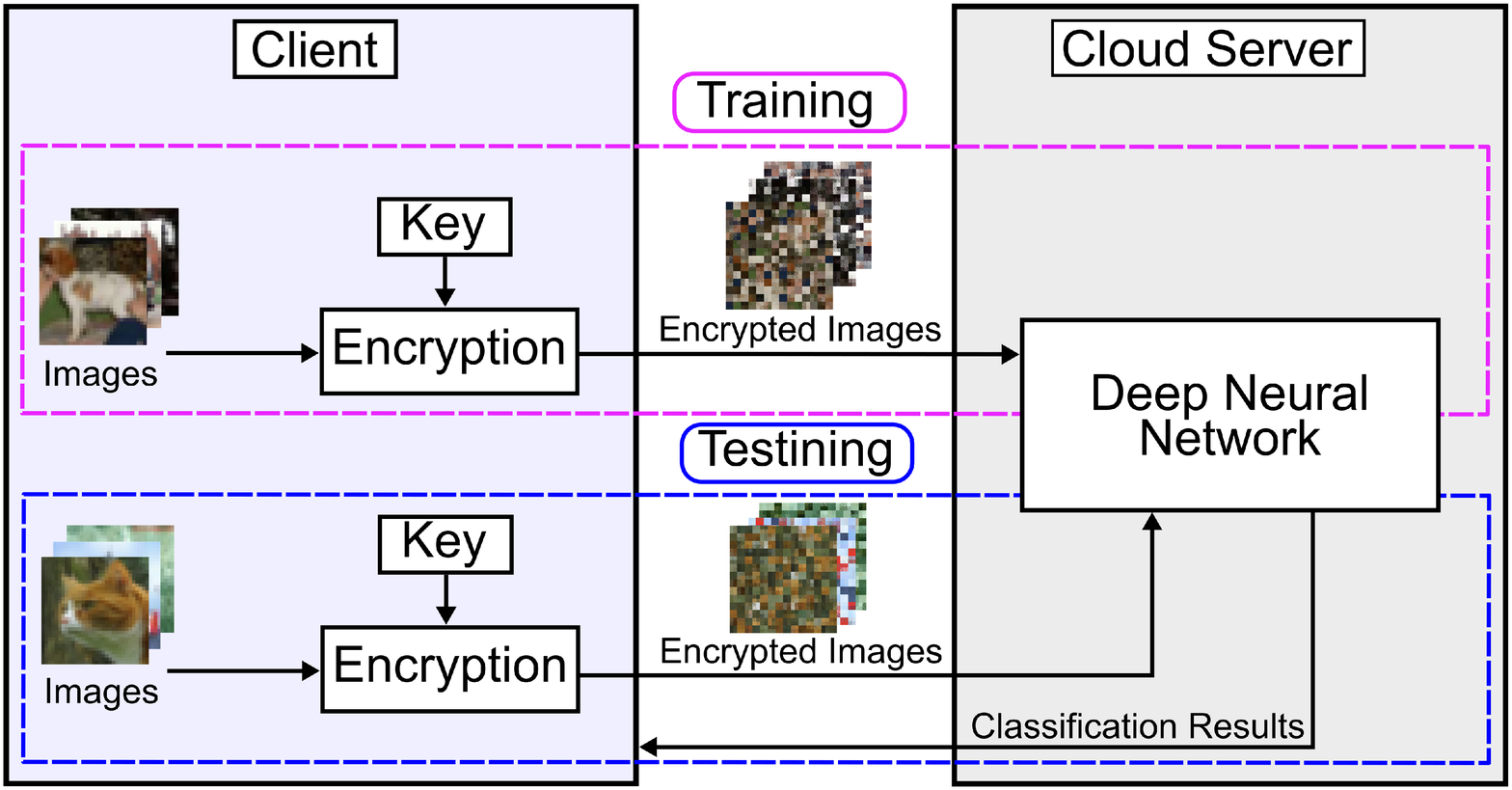}
	\caption{Image classification using deep neural network}
	\label{fig:scheme}
	\end{center}
\end{figure}
\begin{figure}[t]
	\centering
	\includegraphics[width = 12cm]{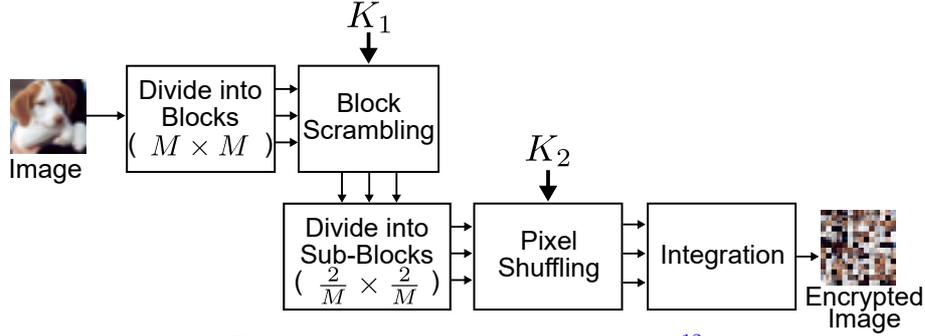}
	\caption{Block-wise image encryption\cite{qi_vit_2022}}
	\label{fig:ex_scheme}
\end{figure}

\section{Preparation}
In this section, the procedure of image encryption for 24-bit RGB color images is described.
Furthermore, the several attacks for this encryption scheme are addressed.
\subsection{Block-wise Image Encryption}
Several block-wise image encryption schemes have been proposed for privacy-preserving DNNs as illustrated in Fig.\ref{fig:scheme}.  
We focus on the encryption scheme\cite{qi_vit_2022}, which combines block-wise image encryption including block and pixel shuffling, and the ViT.
Figure \ref{fig:ex_scheme} indicates the procedure of block-wise image encryption.
The procedure of performing this encryption scheme to generate an encrypted image $I'$ from an image $I$ is given as follows.
\begin{enumerate}
    \item An image $I$ with $X \times Y$ pixels is divided into square blocks $B = \{B_{1},...,B_{i},...,B_{n}\}$, $i \in \{1,...,n\}$ with $M \times M$ pixels, where $n$ is the number of divided blocks calculated by 
    \begin{equation}
        \label{eq:blocknum}
        n = \lfloor \frac{X}{M} \rfloor \times \lfloor \frac{Y}{M} \rfloor.
    \end{equation}
    \item Permute the divided blocks by using a secret key $K_{1}$, where $K_{1}$ is commonly used for all color components. Accordingly, the scrambled blocks $B' = \{B'_{1},...,B'_{i},...,B'_{n}\}$ are generated.
    \item Divide each scrambled block $B'_{i}$ into four non-overlapped square sub-blocks $S_{ij}, j\in \{UL,UR,LL,LR\}$ with $\frac{M}{2} \times \frac{M}{2}$ pixels, where $S_{iUL}$ is defined as upper left position of $i$th blocks, $S_{iUR}$ as upper right, $S_{iLL}$ as lower left, and $S_{iLR}$ as lower right.
    Thereby, scrambled blocks divided into sub-blocks $S = \{S_{1j},...,S_{ij},...,S_{nj}\}$ are generated.
    The number of sub-blocks $m$ is described as
    \begin{equation}
        \label{eq:subblocknum}
        m = 4n.
    \end{equation}
    \item Shuffle the pixel position within a sub-block by using a secret key $K_{2}$ to generate pixel shuffled sub-blocks $S' = \{S'_{1j},...,S'_{ij},...,S'_{nj}\}$, where $K_{2}$ is commonly used for all sub-blocks asnd color components.
    By this operation, each scrambled block consists of four encrypted sub-blocks $S'_{i} = \{S'_{iUL},S'_{iUR},S'_{iLL},S'_{iLR}\}$.
    \item Merge all blocks to generate an encrypted image $I'$
\end{enumerate}
\begin{figure}[!t]
    \captionsetup[subfigure]{justification=centering}
    \centering
    \hspace{1mm}\subfloat[Original]{\includegraphics[clip, width=2.7cm]{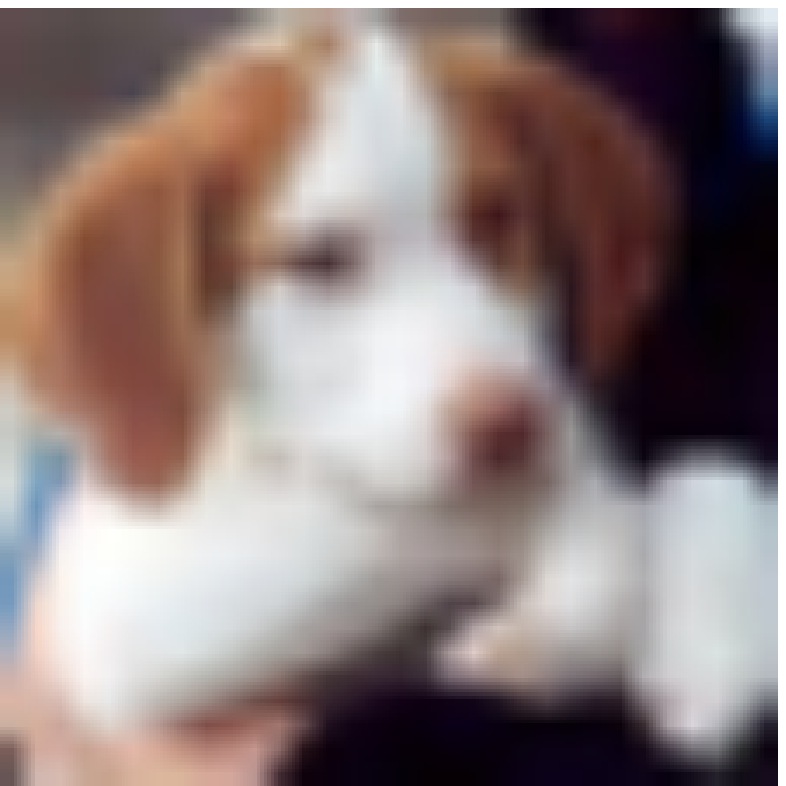}
    \label{fig:00004}}
    \hfil
    \hspace{1mm}\subfloat[Encrypted\cite{qi_vit_2022}]{\includegraphics[clip, width=2.7cm]{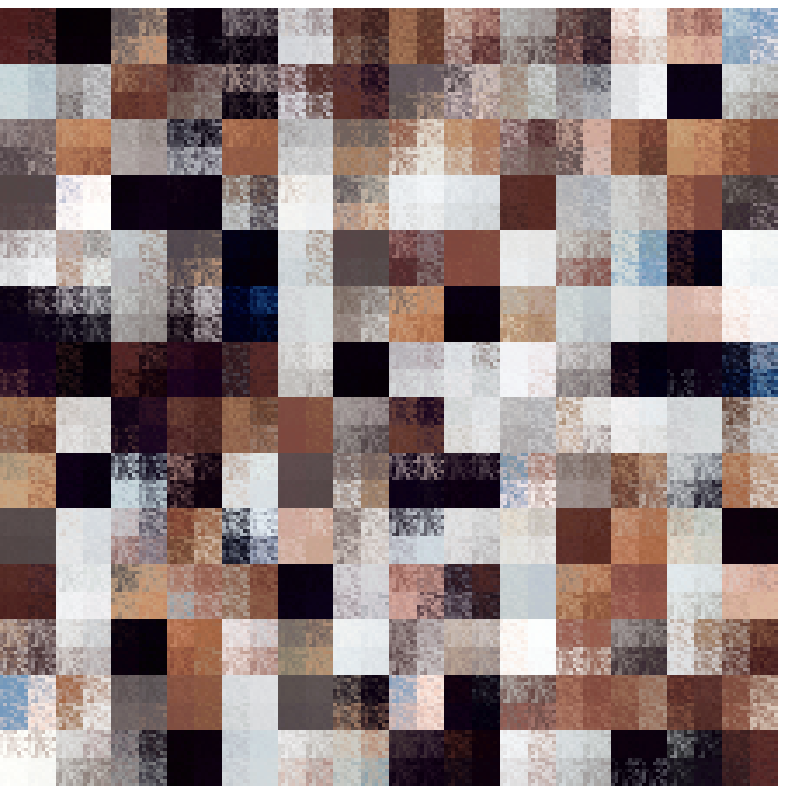}
    \label{fig:00004_enc}}
    \caption{Examples of encrypted images by using the block-wise image encryption($X \times Y = 224 \times 224, M = 16$)}
    \label{fig:organdencs}
\end{figure}
\begin{figure}[t]
	\centering
	\includegraphics[width = 10cm]{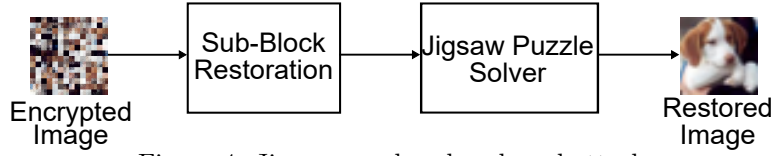}
	\caption{Jigsaw puzzle solver-based attack}
	\label{fig:pro_scheme}
\end{figure}
\begin{figure}[!t]
	\centering
	\includegraphics[width = 7cm]{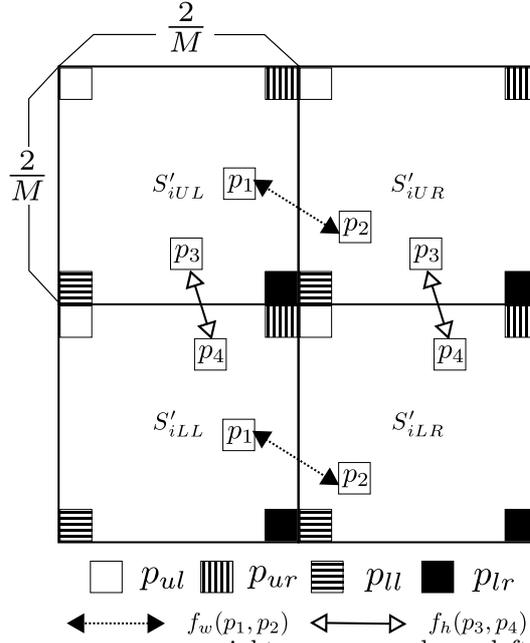}
	\caption{Positions of upper left corner ${p_{ul}}$, upper right corner ${p_{ur}}$, lower left corner ${p_{ll}}$, and lower right corner ${p_{lr}}$}
	\label{fig:position}
\end{figure}
An example of an encrypted image is shown in Fig.\ref{fig:organdencs}\subref{fig:00004_enc}; Fig.\ref{fig:organdencs}\subref{fig:00004} shows the original one.
In this paper, $M = 16$ is utilized as well as in \cite{qi_vit_2022}.
\subsection{Cipher-text only Attacks}
Several cipher-text only attacks (COAs) have been proposed to restore encrypted images.
Although several attacks have succeeded in restoring the encrypted images, it has been known that the use of both block scrambling and pixel shuffling enhances robustness against COAs\cite{tanaka_AAAI}.
\par
On the other, the encrypted image including scrambled blocks is restored by using the jigsaw puzzle solver attack, which regards the blocks of an encrypted image as pieces of a jigsaw puzzle.
However, the block-wise image encryption\cite{qi_vit_2022} becomes robust against the jigsaw puzzle solver attack because of combining block scrambling and pixel shuffling.
Thus, we propose the jigsaw puzzle solver-based attack, which enables us to restore encrypted images including permuted blocks and pixels.

\begin{figure*}[t]
	\centering
	\includegraphics[width = \linewidth]{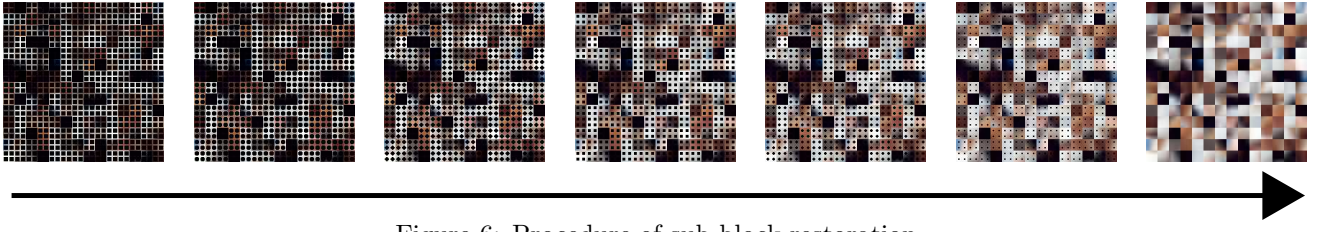}
	\caption{Procedure of sub-block restoration}
	\label{fig:image_rec}
\end{figure*}

\section{Proposed Scheme}
In this paper, the jigsaw puzzle solver-based attack, which utilizes the correlation between the blocks for restoring, is proposed.
The proposed attack consists of two steps as illustrated in Fig.\ref{fig:pro_scheme}: the first step is sub-block restoration, which aims to solve pixel shuffling of the encrypted image; the second step is a jigsaw puzzle solver attack to assemble the scrambled image.
\subsection{Sub-block Restoration}
The purpose of sub-block restoration is to solve the pixel shuffling in sub-block by utilizing the pixels of edge in sub-block, namely generate non-shuffled pixel block $\hat{S}_{ij}$ from $S'_{ij}$.
\par
Firstly, the positions of upper left corner ${p_{ul}}$, upper right corner ${p_{ur}}$, lower left corner ${p_{ll}}$, and lower right corner ${p_{lr}}$ in $\hat{S}_{ij}$ are determined as illustrated in Fig.\ref{fig:position}.
Given for the four different pixel positions $p_{1}=(x_{1}, y_{1})$, $p_{2}=(x_{2}, y_{2})$, $p_{3}=(x_{3}, y_{3})$, $p_{4}=(x_{4}, y_{4})$, $p_{1} \neq p_{2} \neq p_{3} \neq p_{4},$ $x_{1}, y_{1}, x_{2}, y_{2}, x_{3}, y_{3}, x_{4}, y_{4}  \in \{1,2,...,\frac{M}{2}\}$ in $S'_{ij}$, the pixel intensities are defined as $S'_{ij}(p_{1},c)$, $S'_{ij}(p_{2},c)$, $S'_{ij}(p_{3},c)$ and $S'_{ij}(p_{4},c)$, $c \in \{R,G,B\}$. Thus, the sum of all mean squared error (MSE) values between left and right sub-blocks is calculated by
\begin{eqnarray}
    \begin{aligned} 
    \label{eq:sum_mse_w}
    f_{w}(p_{1},p_{2}) = \sum_{c}^{}\sum_{i=1}^{n} ({S'_{iUL}(p_{1},c) - S'_{iUR}(p_{2},c)})^2
    \\ +({S'_{iLL}(p_{1},c) - S'_{iLR}(p_{2},c)})^2.
    \end{aligned} 
\end{eqnarray}
On the other hand, the sum of all MSE values between upper and lower sub-blocks is calculated as 
\begin{eqnarray}
    \begin{aligned} 
    \label{eq:sum_mse_h}
    f_{h}(p_{3},p_{4}) = \sum_{c}^{}\sum_{i=1}^{n} ({S'_{iUL}(p_{3},c) - S'_{iLL}(p_{4},c)})^2
    \\ +({S'_{iUR}(p_{3},c) - S'_{iLR}(p_{4},c)})^2.
    \end{aligned}
\end{eqnarray}
Using Eq.(\ref{eq:sum_mse_w}) and (\ref{eq:sum_mse_h}), ${p_{ul}}$, ${p_{ur}}$, ${p_{ll}}$  and ${p_{lr}}$ are given as
\begin{eqnarray}
    \begin{aligned}
    \label{eq:sum_mse_corner}
    p_{lr},p_{ll},p_{ur},p_{ul}
    = \argmin_{p_{1},p_{2},p_{3},p_{4}}\{f_{w}(p_{1},p_{2}) + f_{h}(p_{1},p_{3}) \\
    + f_{w}(p_{3},p_{4}) + f_{h}(p_{2},p_{4})\}.
    \end{aligned}
\end{eqnarray}
\par
Next, the positions of restored right edge $p_{r} = \{p_{r1},...,p_{rk},..., p_{r\frac{M}{2}-2}\}$ and left edge $p_{l} = \{p_{l1},...,p_{lk},..., p_{l\frac{M}{2}-2}\}$ are calculated as
\begin{eqnarray}
    \begin{aligned} 
    \label{eq:sum_mse_w_arg}
        p_{rk}, p_{lk} = \argmin_{p_{1},p_{2}}{f_{w}(p_{1},p_{2})}.
    \end{aligned} 
\end{eqnarray}
\par
As well as $p_{rk}$ and $p_{lk}$, the positions of restored upper edge $p_{u} = \{p_{u1},...,p_{uk},..., p_{u\frac{M}{2}-2}\}$ and lower edge $p_{dy} = \{p_{d1},...,p_{dk},..., p_{d\frac{M}{2}-2}\}$ are defined as
\begin{eqnarray}
    \begin{aligned} 
    \label{eq:sum_mse_h_arg}
        p_{uk}, p_{dk} = \argmin_{p_{3},p_{4}}{f_{h}(p_{3},p_{4})}.
    \end{aligned} 
\end{eqnarray}
\par
The remaining positions in $\hat{S}_{ij}$ are determined by minimizing the MSE of surrounding pixels to generate the restored image $\hat{I}$ as illustrated in Fig.\ref{fig:image_rec}, where Fig.\ref{fig:organdencs}\subref{fig:00004_enc} is the encrypted one before performing the sub-block restoration.

\subsection{Jigsaw Puzzle Solver Attack}
After solving the pixel shuffling encryption by using the sub-block restoration, the jigsaw puzzle solver is applied to $\hat{I}$ to generate restored image $\hat{I}'$.
It has been known that the encrypted image including only shuffled blocks with $14 \times 14$ pixels can be restored easily\cite{CHUMAN_IEICE}.
Therefore, the encrypted image by using the block-wise image encryption with $M = 16$ can be restored when the sub-block restoration performs well.

\section{Experiments}
\subsection{Experimental Conditions}
In this section, the security of the block-wise image encryption was evaluated by using the conventional and proposed attack.
In addition to the use of the proposed attack, encrypted images were restored by using the conventional one, which employs only the jigsaw puzzle solver, to demonstrate effectiveness of the sub-block restoration.
A jigsaw puzzle solver using genetic algorithms was utilized to restore encrypted images\cite{Sholomon_2016_GPEM}.
\par
We employed the 10,000 testing images with 32 $\times$ 32 pixels from CIFAR-10 dataset.
Before performing the block-wise image encryption, each image is resized to 224 $\times$ 224 pixels to fit the same patch size of pretrained model such as ViT-B/16 and ViT-L/16.
The average of the structural similarity index measure (SSIM) values between an original image and the restored one were calculated. 

\subsection{Experimental Results}
Table \ref{table:result} shows the result of security evaluation of block-wise image encryption against the conventional and proposed attack.
Although the SSIM value of restored images by using the conventional attack was low as 0.35, the proposed scheme enabled to restore encrypted images mostly as the value of SSIM was 0.97.
As illustrated in Fig.\ref{figure:ex_result}, the proposed attack succeeded in reconstructing the encrypted image more clearly than the conventional attack.

\begin{table}[!t]
    \centering
    \caption{Security evaluation of block-wise image encryption}
    \label{table:result}
    \scalebox{1.0}{
    \begin{tabular}{|c|c|}
    \hline
    Attack       & SSIM \\ \hline
    Conventional & 0.35 \\ \hline
    Proposed  & 0.97 \\ \hline

    \end{tabular}
    }
\end{table}
\section{Conclusion}
In this paper, we proposed the jigsaw puzzle solver-based attack on block-wise image encryption for privacy-preserving DNNs.
In experiments, robustness of the encrypted images by using the block-wise image encryption, which include scrambled blocks and shuffled pixels, was evaluated by using the conventional and proposed attacks. 
Experimental results showed that the block-wise image encryption was robustness against the conventional attack.
On the other hand, the effectiveness of the proposed attack, which consists of the sub-block restoration and jigsaw puzzle solver was confirmed. 
\newcommand{\orione}{\includegraphics[width=4.5em]{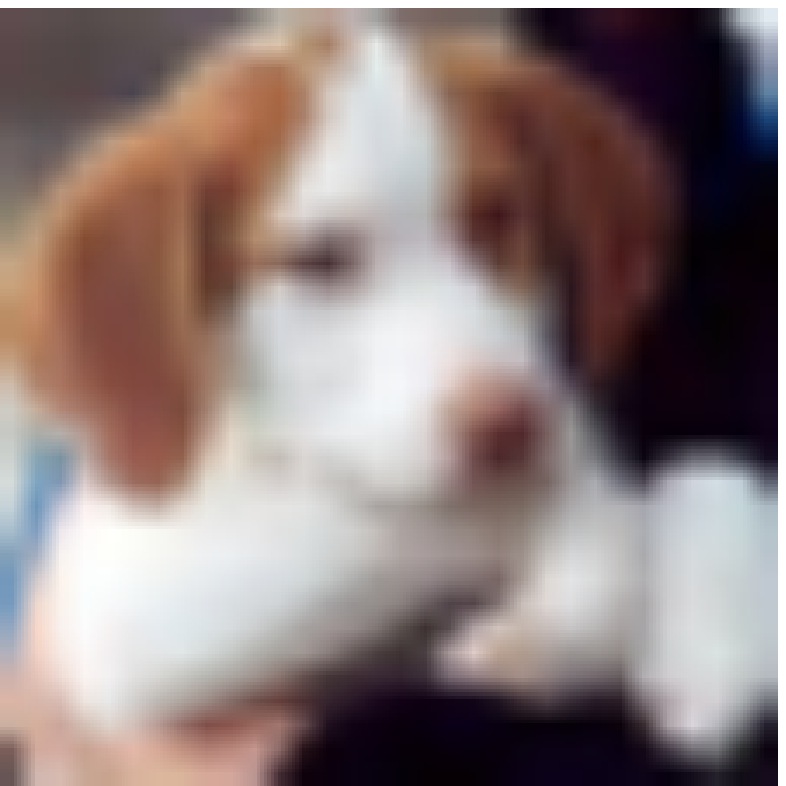}}
\newcommand{\oritwo}{\includegraphics[width=4.5em]{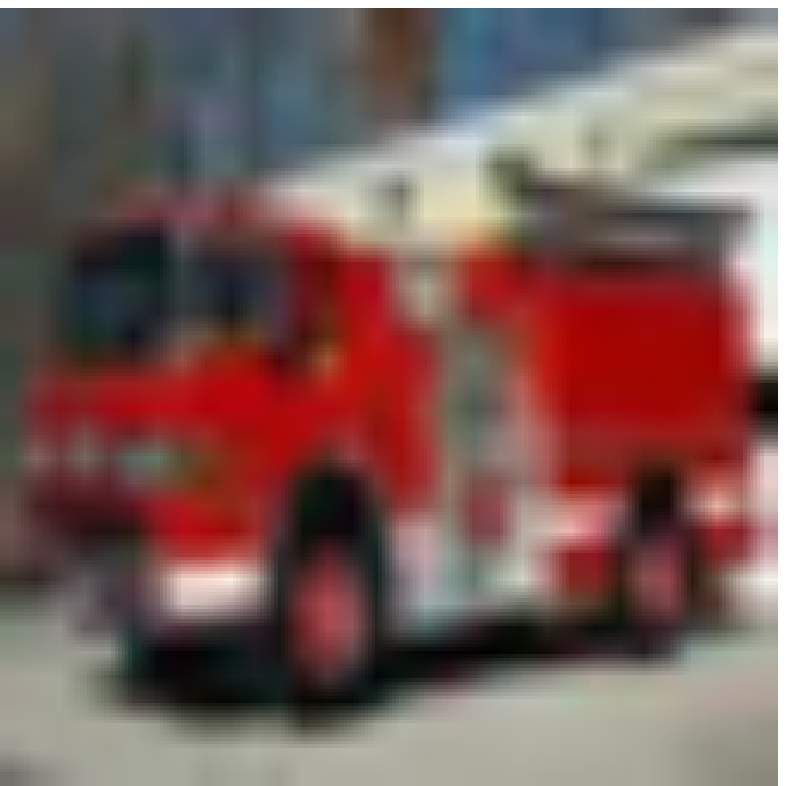}}
\newcommand{\orithree}{\includegraphics[width=4.5em]{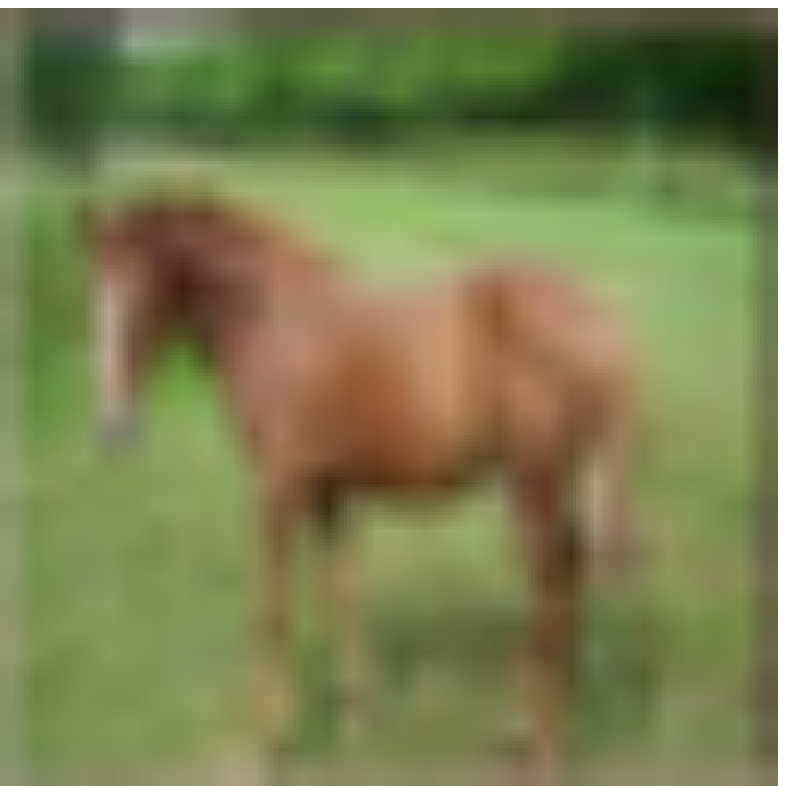}}
\newcommand{\orifour}{\includegraphics[width=4.5em]{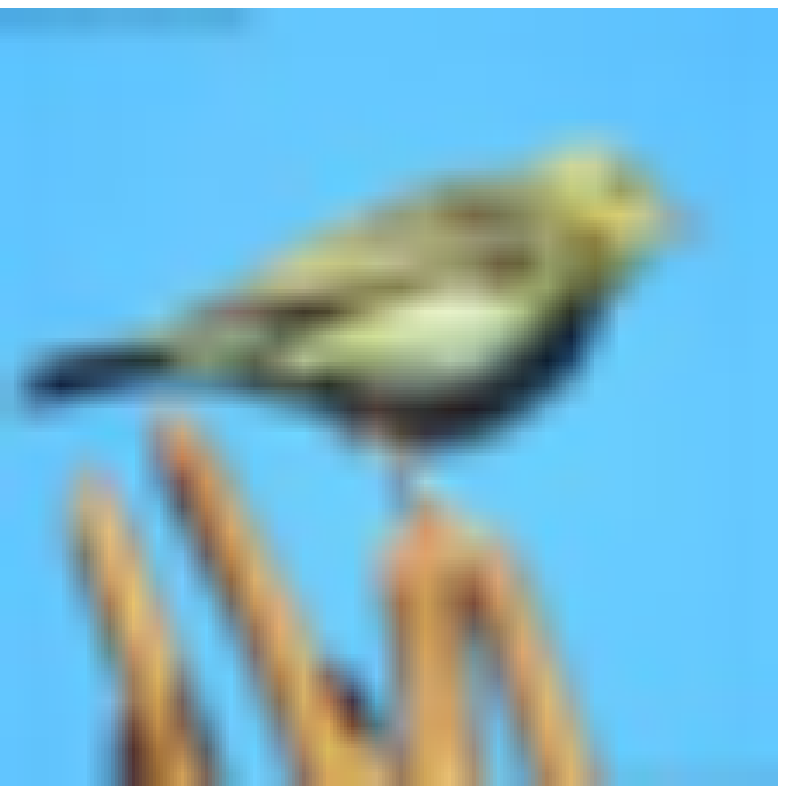}}
\newcommand{\orifive}{\includegraphics[width=4.5em]{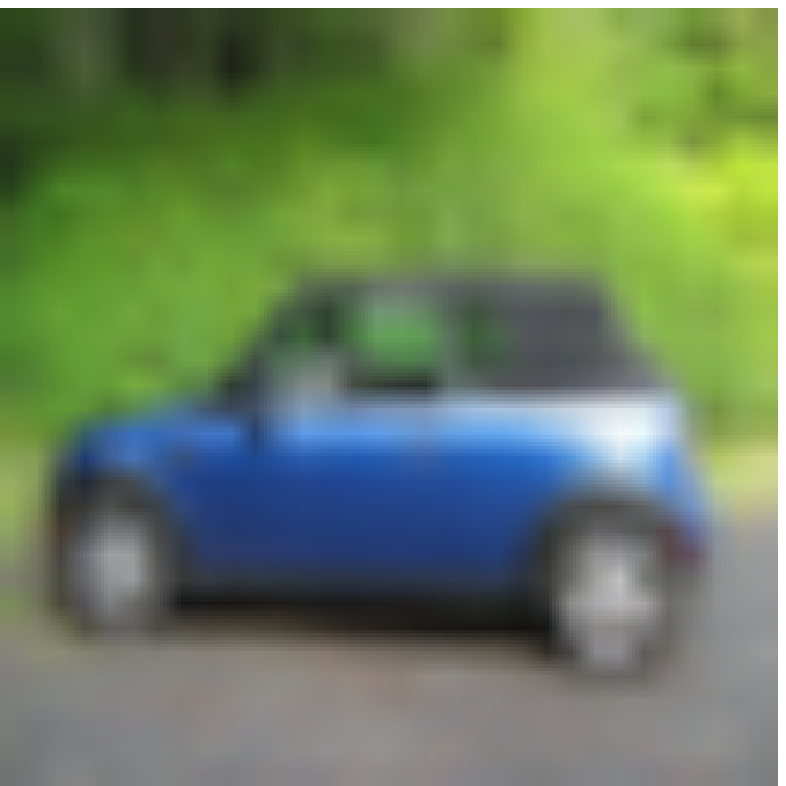}}
\newcommand{\orisix}{\includegraphics[width=4.5em]{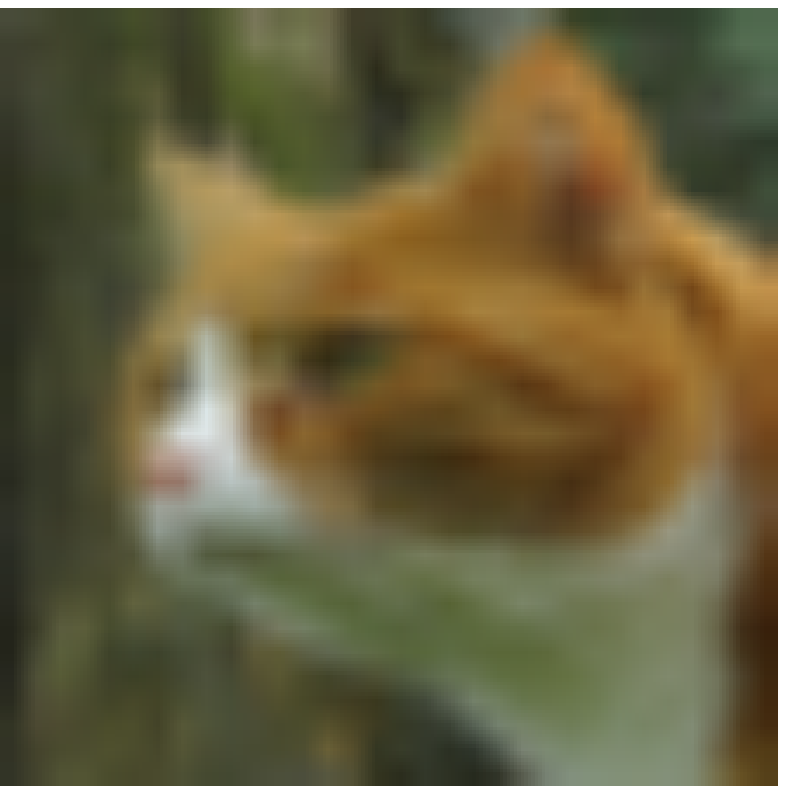}}
\newcommand{\oriseven}{\includegraphics[width=4.5em]{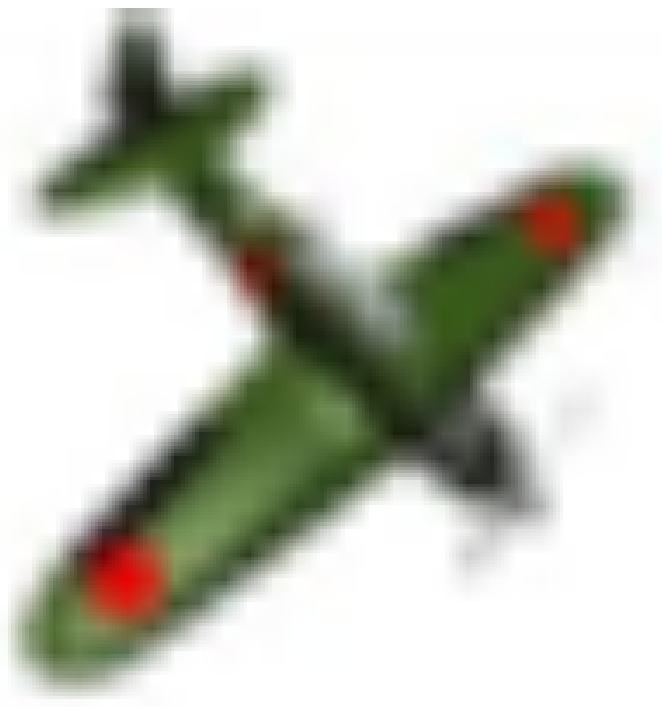}}

\newcommand{\encone}{\includegraphics[width=4.5em]{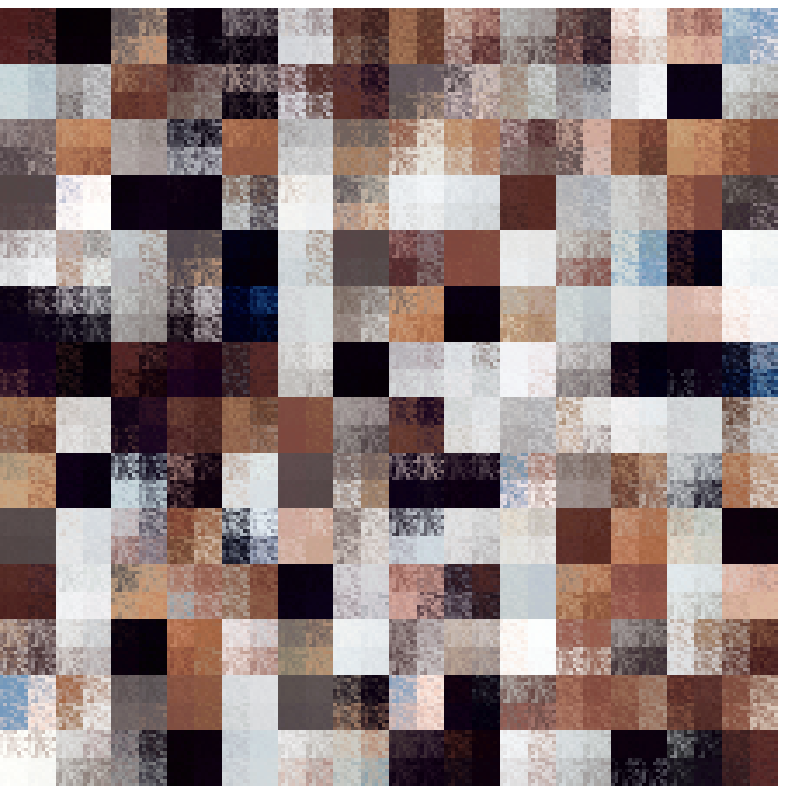}}
\newcommand{\enctwo}{\includegraphics[width=4.5em]{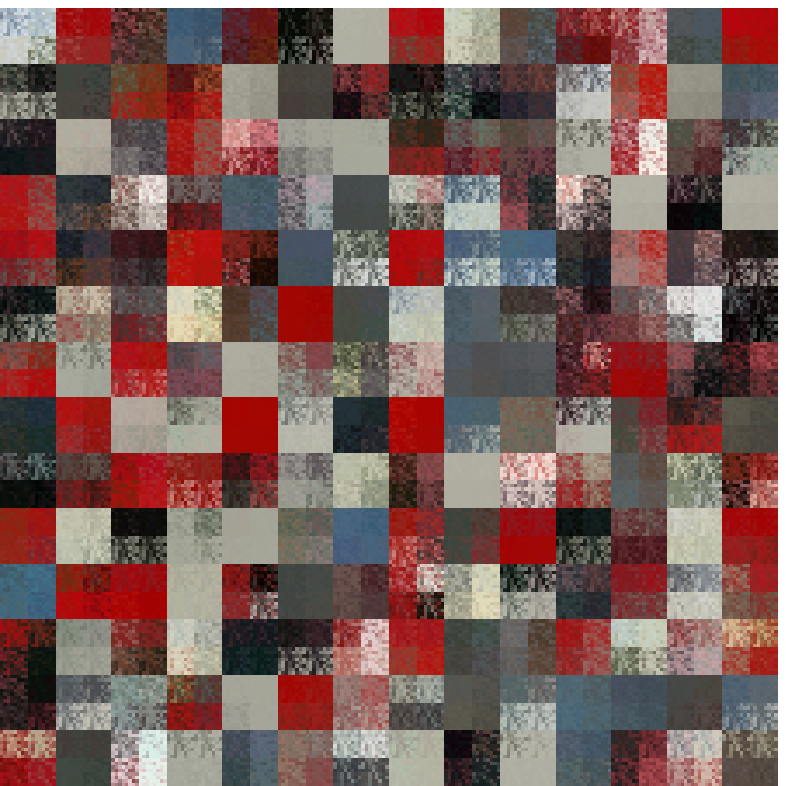}}
\newcommand{\encthree}{\includegraphics[width=4.5em]{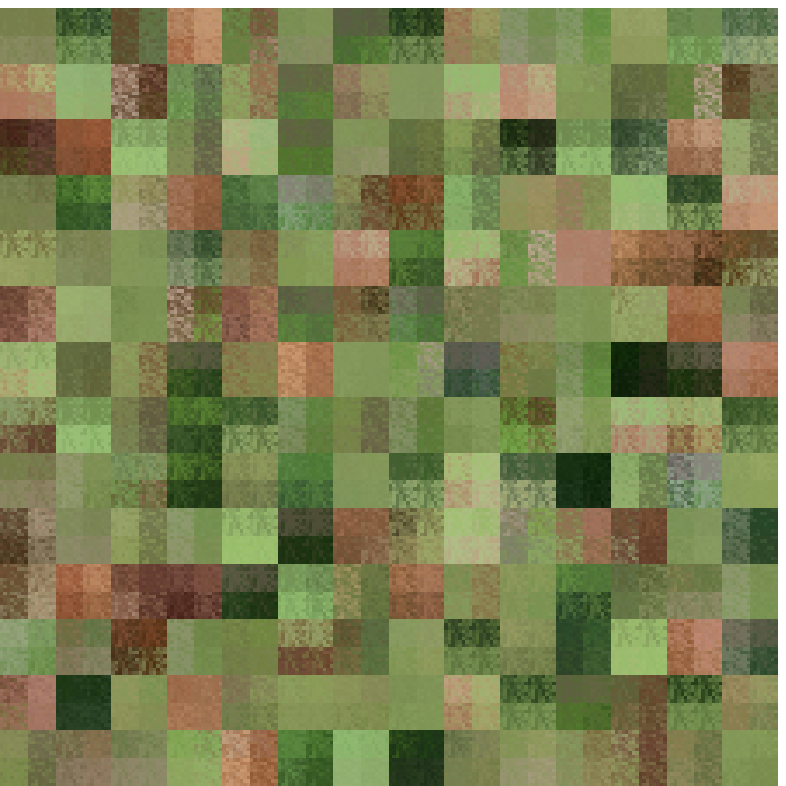}}
\newcommand{\encfour}{\includegraphics[width=4.5em]{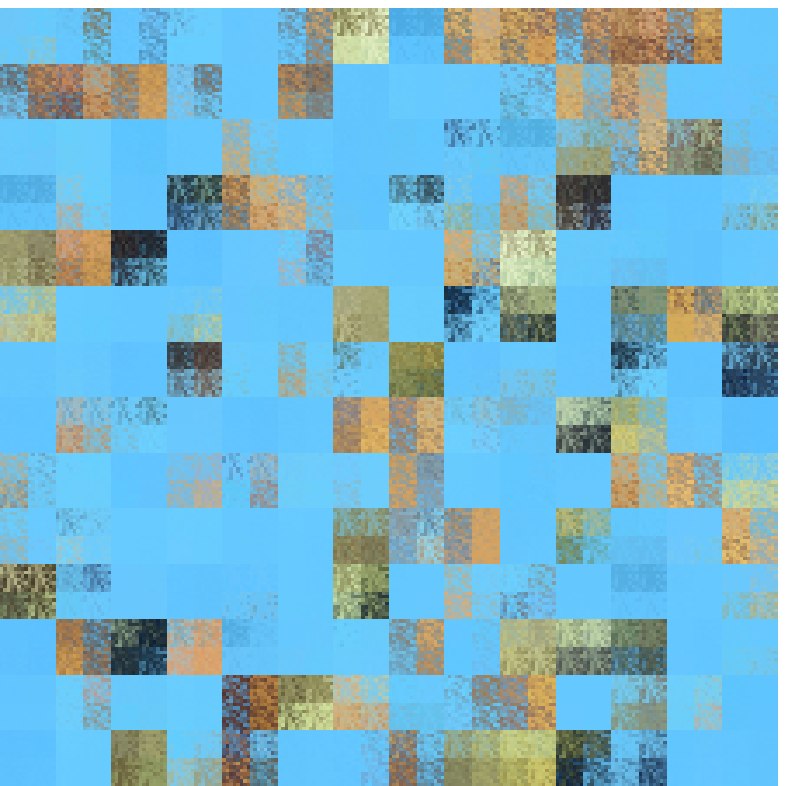}}
\newcommand{\encfive}{\includegraphics[width=4.5em]{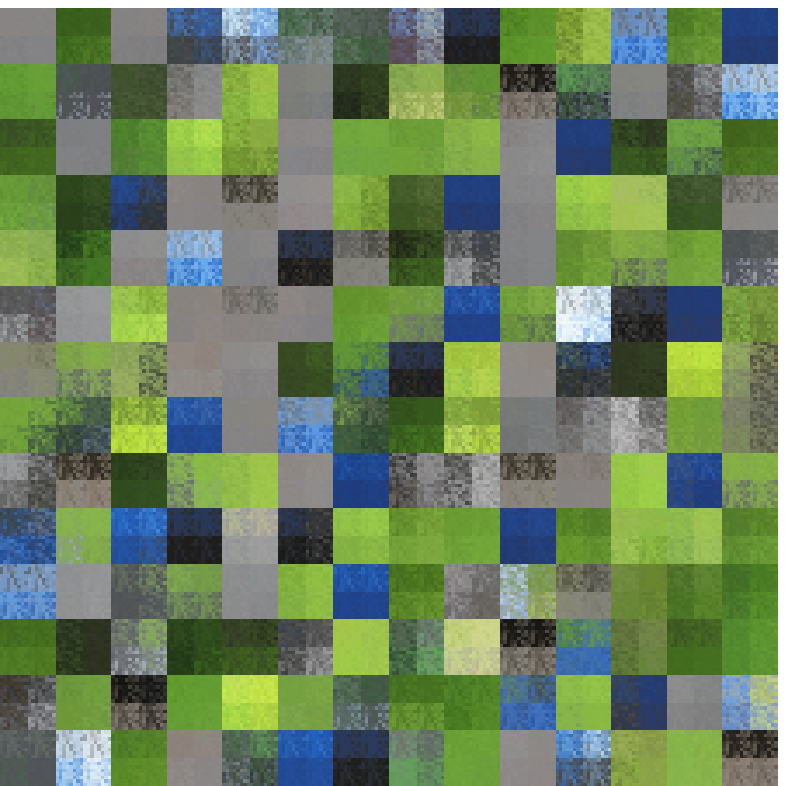}}
\newcommand{\encsix}{\includegraphics[width=4.5em]{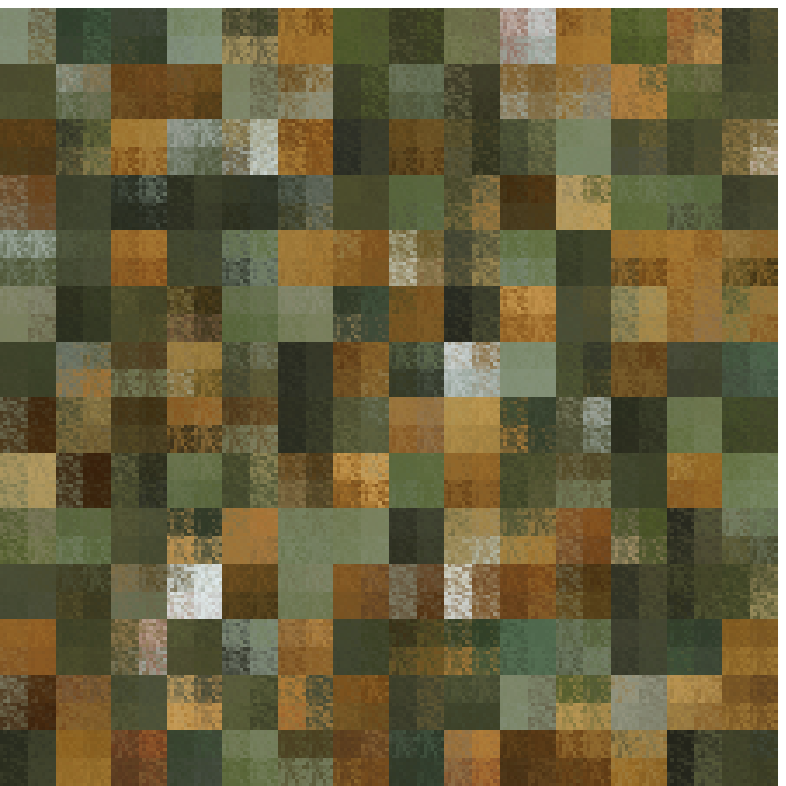}}
\newcommand{\encseven}{\includegraphics[width=4.5em]{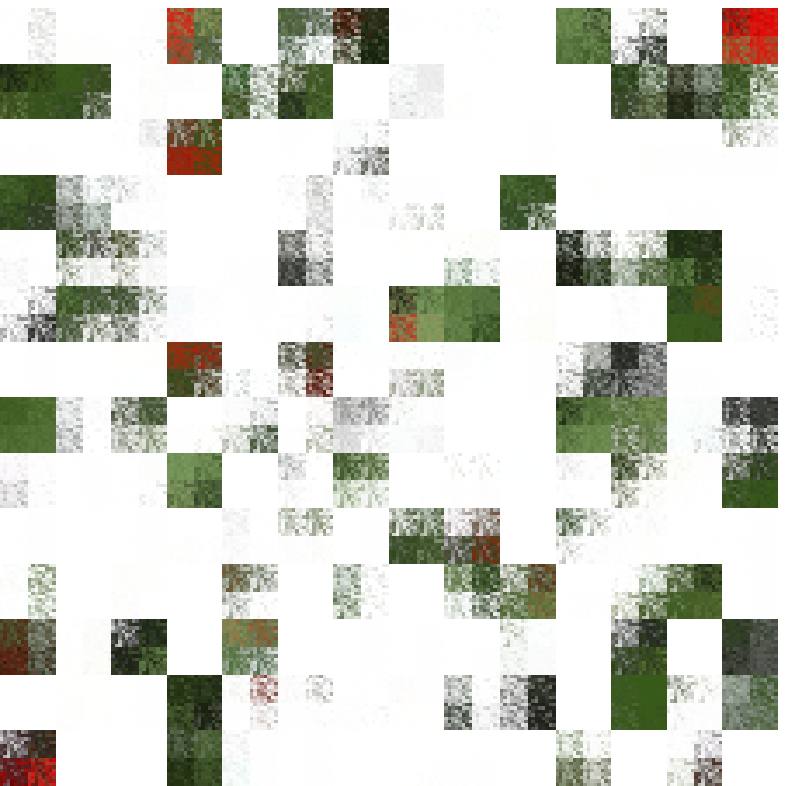}}

\newcommand{\convone}{\includegraphics[width=4.5em]{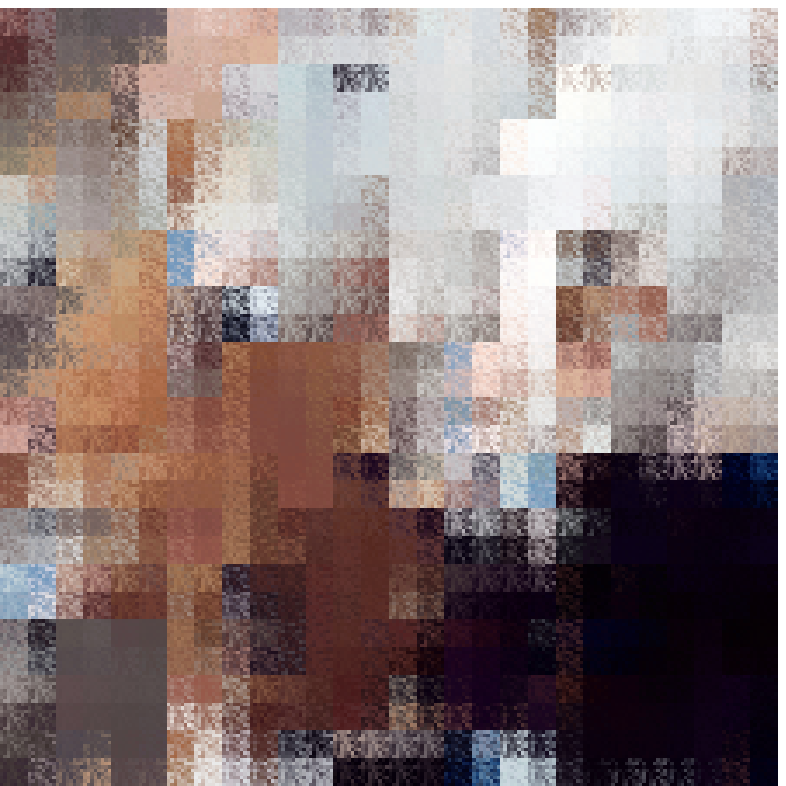}}
\newcommand{\convtwo}{\includegraphics[width=4.5em]{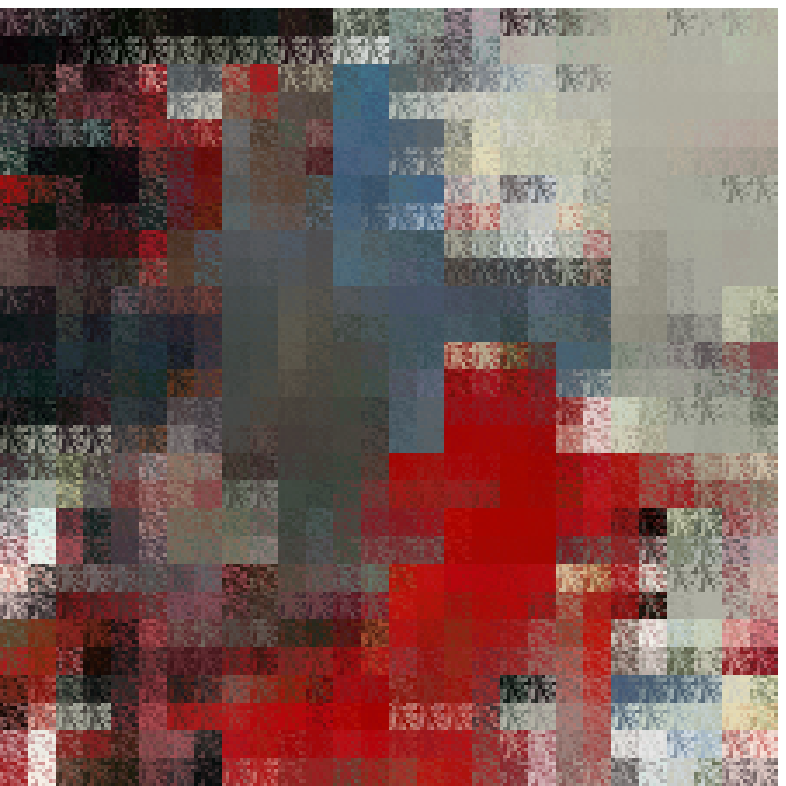}}
\newcommand{\convthree}{\includegraphics[width=4.5em]{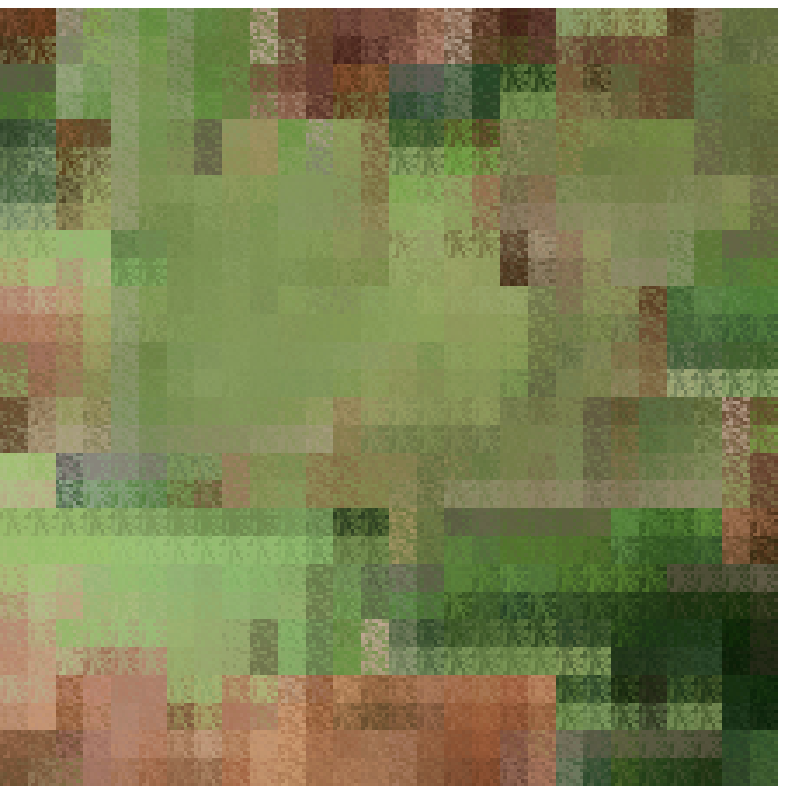}}
\newcommand{\convfour}{\includegraphics[width=4.5em]{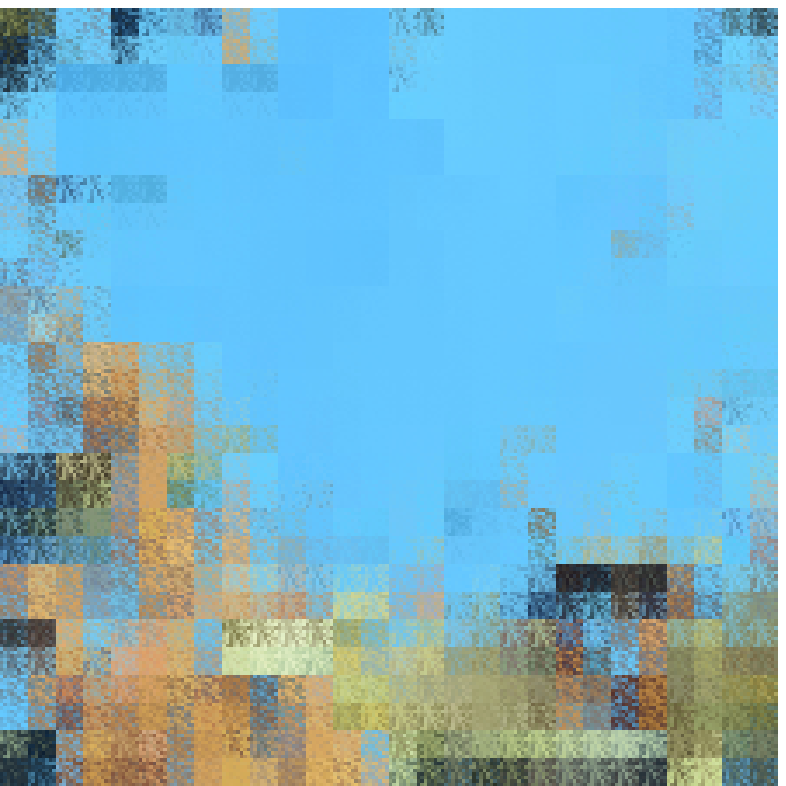}}
\newcommand{\convfive}{\includegraphics[width=4.5em]{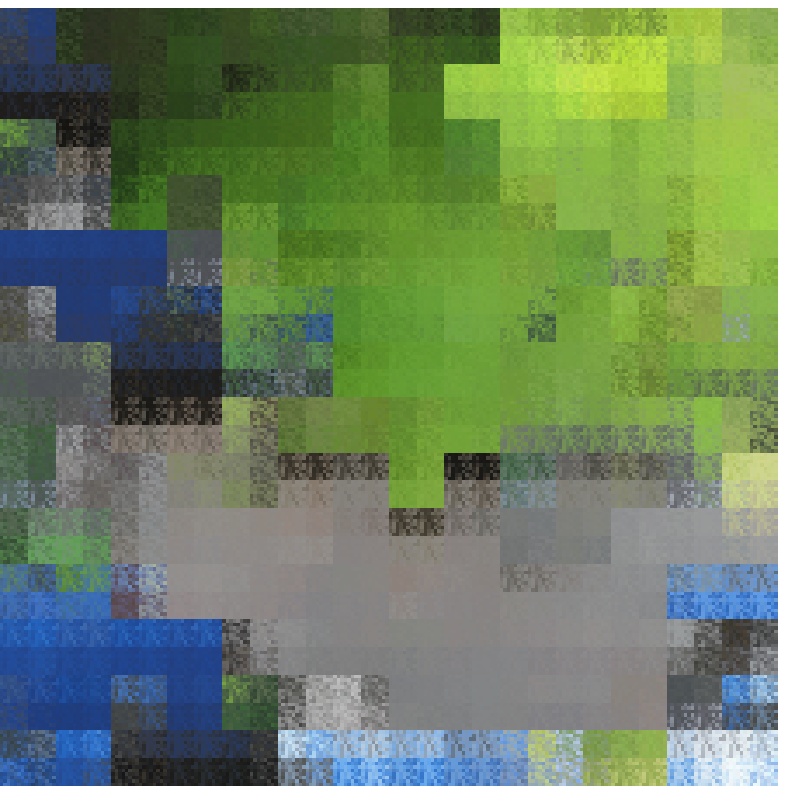}}
\newcommand{\convsix}{\includegraphics[width=4.5em]{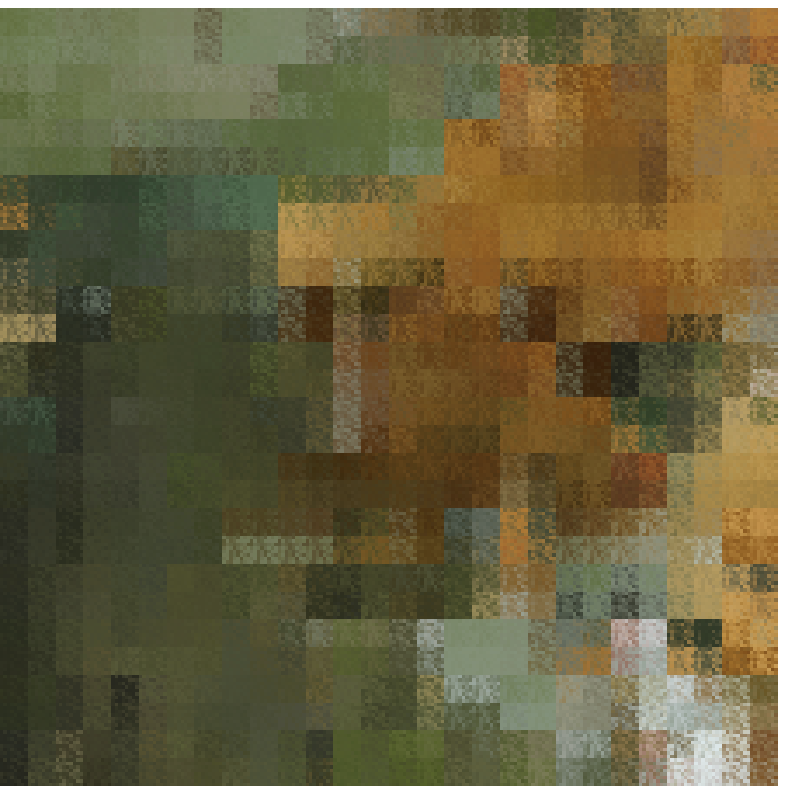}}
\newcommand{\convseven}{\includegraphics[width=4.5em]{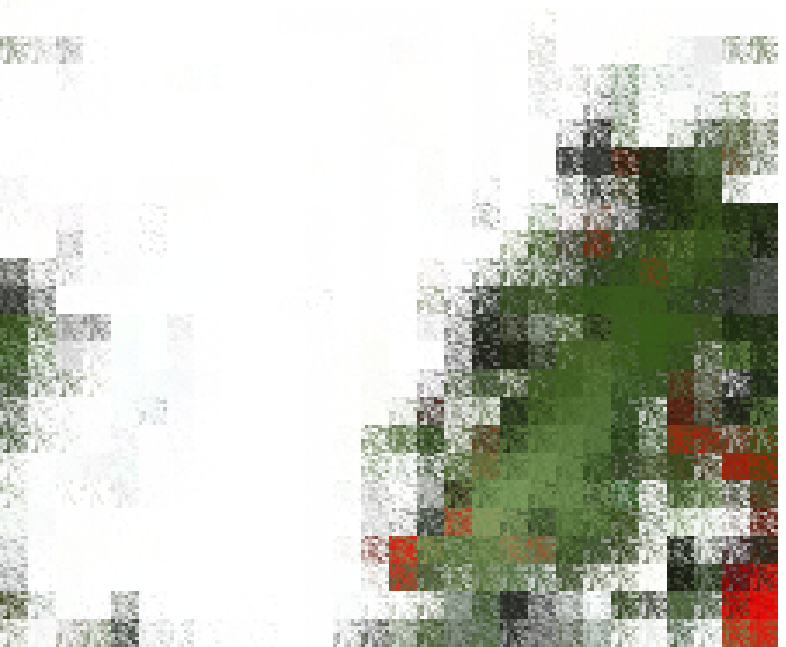}}

\newcommand{\propone}{\includegraphics[width=4.5em]{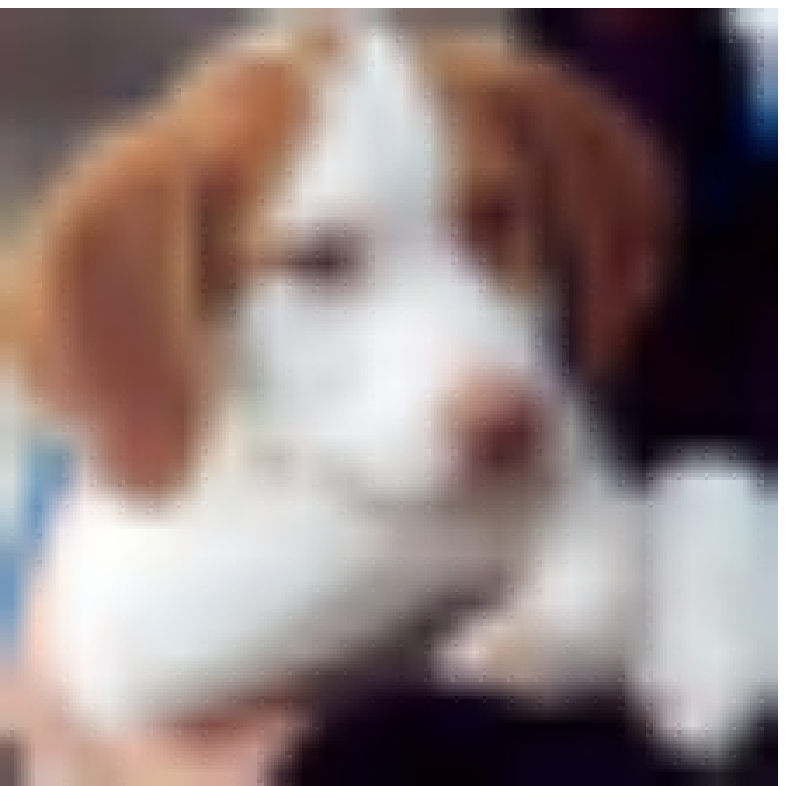}}
\newcommand{\proptwo}{\includegraphics[width=4.5em]{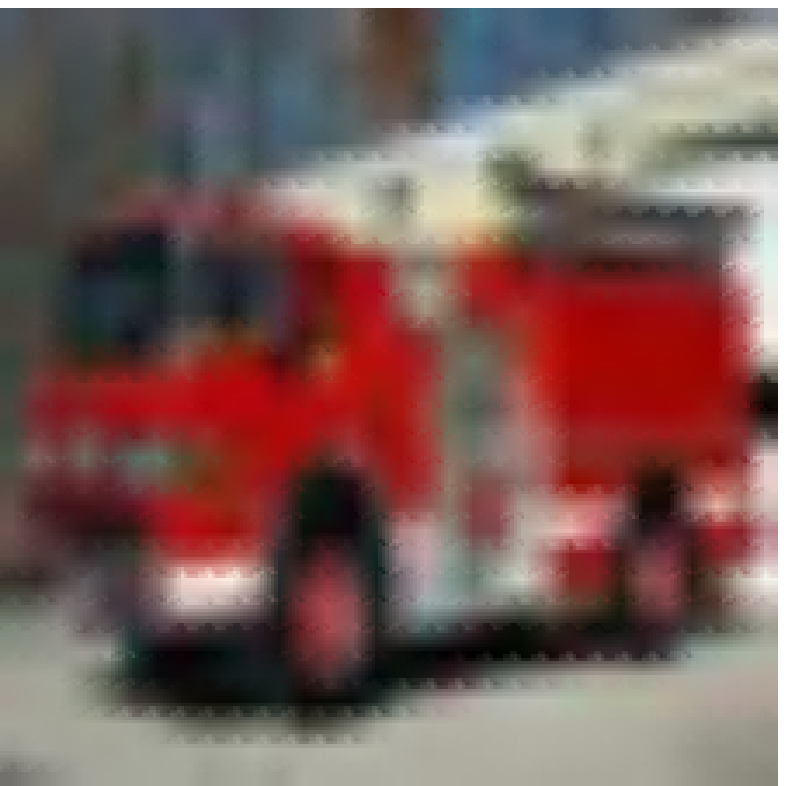}}
\newcommand{\propthree}{\includegraphics[width=4.5em]{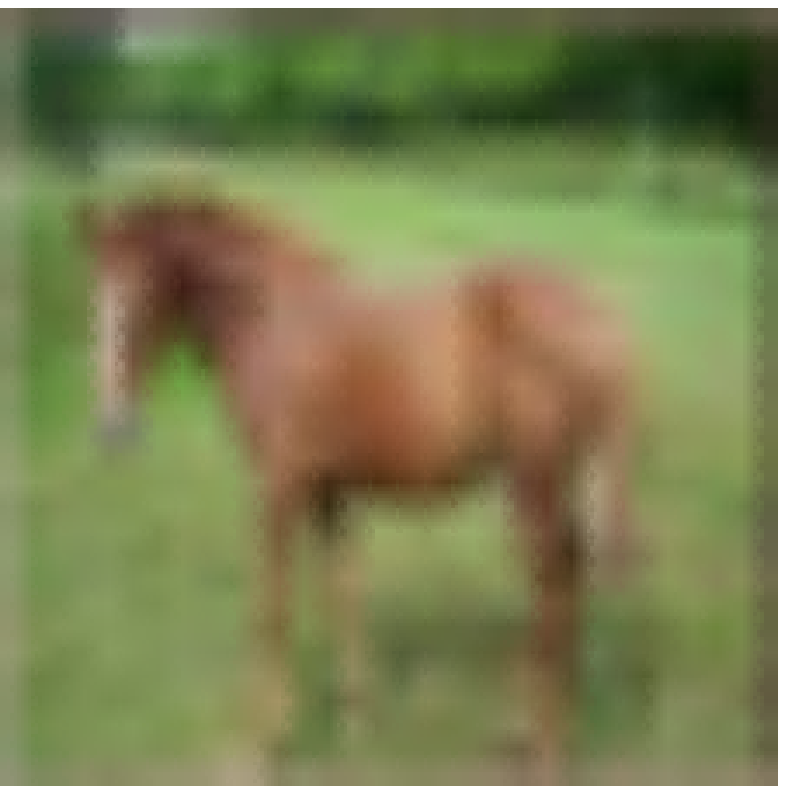}}
\newcommand{\propfour}{\includegraphics[width=4.5em]{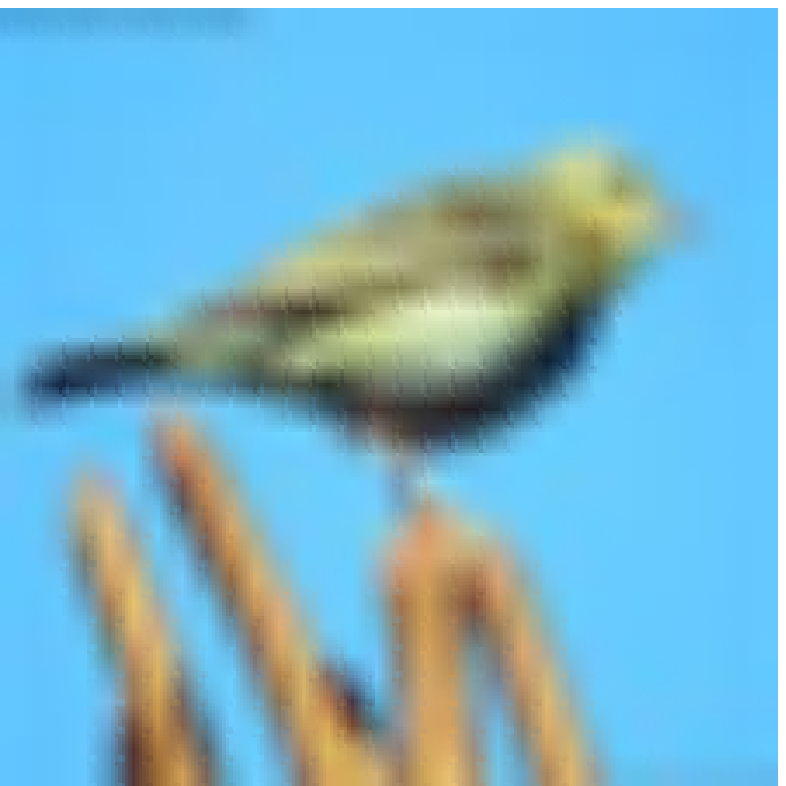}}
\newcommand{\propfive}{\includegraphics[width=4.5em]{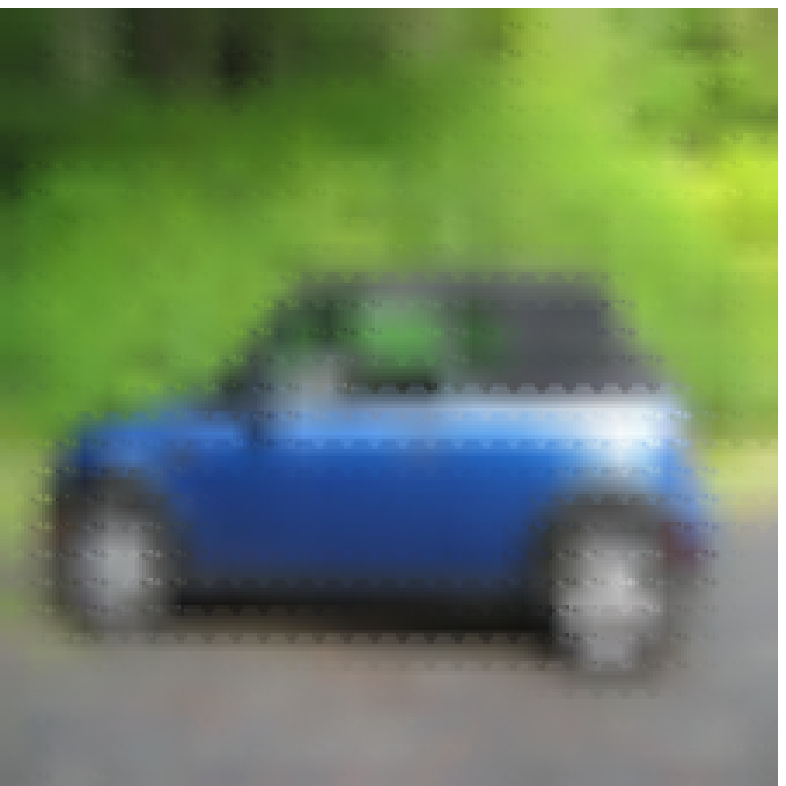}}
\newcommand{\propsix}{\includegraphics[width=4.5em]{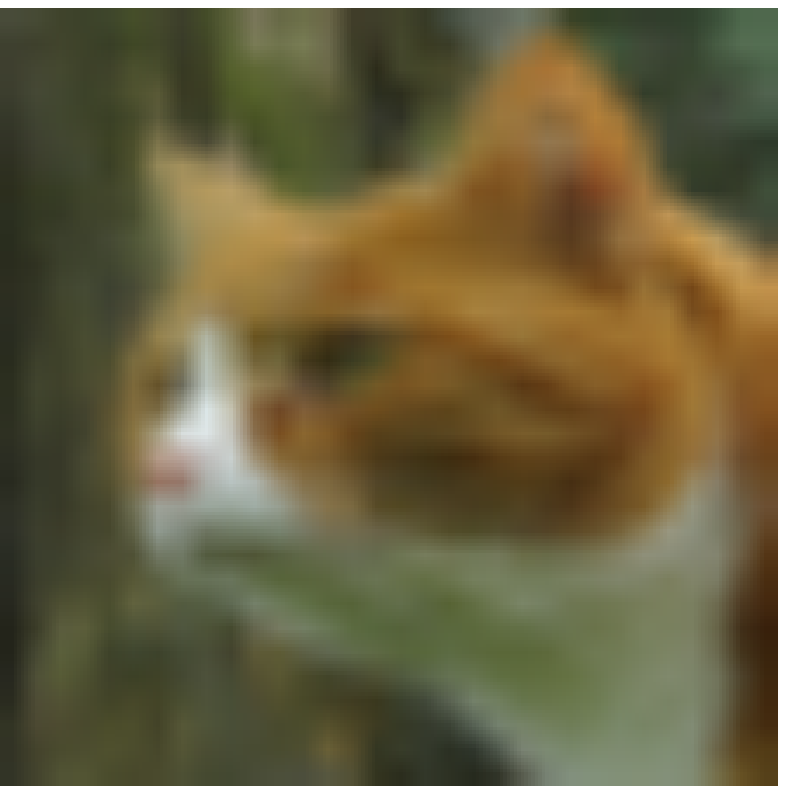}}
\newcommand{\propseven}{\includegraphics[width=4.5em]{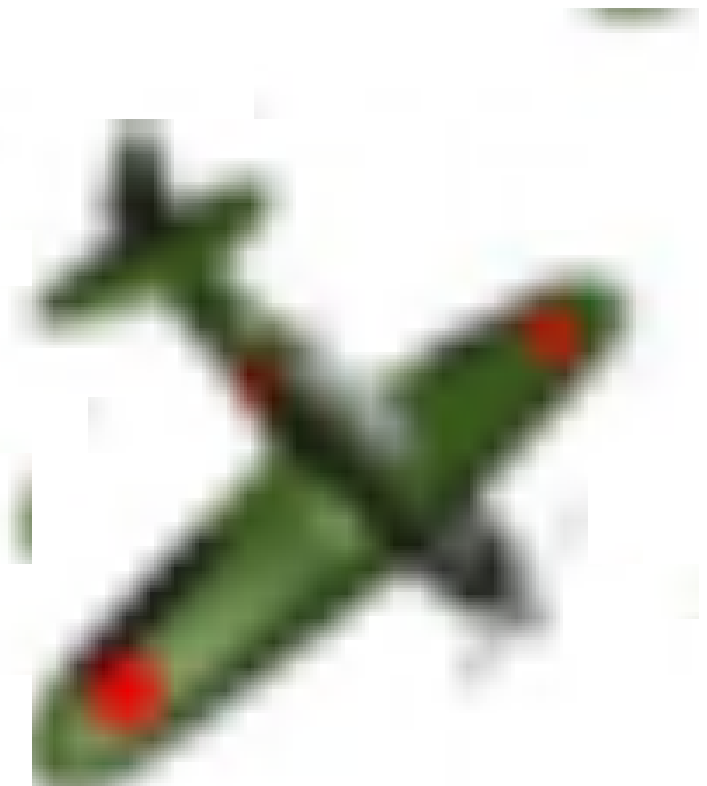}}

\newcolumntype{C}{>{\centering\arraybackslash}m{4.5em}}

\begin{figure*}[t]
	\centering
	\begin{tabular}{l*7{C}@{}}
	Original & \orione & \oritwo & \orithree & \orifour & \orifive & \orisix & \oriseven\\ 
	Encrypted\cite{qi_vit_2022} & \encone & \enctwo & \encthree & \encfour & \encfive & \encsix & \encseven\\ 

	Conventional attack\cite{Sholomon_2016_GPEM} & \convone & \convtwo & \convthree & \convfour & \convfive & \convsix & \convseven\\
	Proposed attack & \propone & \proptwo & \propthree & \propfour & \propfive & \propsix & \propseven\\ 
	\end{tabular}
    \caption{Examples of restored images by using the conventional and proposed attack}
	\label{figure:ex_result}
\end{figure*}
\acknowledgments 
This study was partially supported by JSPS KAKENHI (Grant Number JP21H01327) and the Support Center for Advanced Telecommunications Technology Research, Foundation (SCAT).

\bibliographystyle{spiebib} 
\bibliography{refs_iwait} 

\end{document}